\documentclass[a4paper,onecolumn,quantumarticle,unpublished]{quantumarticle}
\pdfoutput=1
\PassOptionsToPackage{hyphens}{url}
\usepackage[usenames,dvipsnames]{xcolor}
\usepackage{graphicx}
\usepackage[ colorlinks = true, 
             linkcolor = PineGreen,
             urlcolor  = violet,
             citecolor = violet,
             anchorcolor = Magenta,
]{hyperref} 
\usepackage{amsmath,amssymb,amsthm,mathtools,mathrsfs}
\usepackage{pifont}
\usepackage{authblk} 
\usepackage{bbm}
\usepackage{tikz}
\usepackage[compat=1.1.0]{tikz-feynman}

\usepackage{dsfont}
\usepackage{enumerate}
\usepackage{tocbasic}

\DeclareTOCStyleEntry[
  beforeskip=0.3em plus 1pt,
  pagenumberformat=\textbf
]{tocline}{section}

\makeatletter
\renewcommand*\env@matrix[1][\arraystretch]{%
  \edef\arraystretch{#1}%
  \hskip -\arraycolsep
  \let\@ifnextchar\new@ifnextchar
  \array{*\c@MaxMatrixCols c}}
\makeatother

\theoremstyle{plain}
\newtheorem{theorem}[equation]{Theorem}
\newtheorem{lemma}[equation]{Lemma}
\newtheorem{proposition}[equation]{Proposition}
\newtheorem{corollary}[equation]{Corollary}
\theoremstyle{definition}
\newtheorem{definition}[equation]{Definition}

\newtheorem{question}[equation]{Question}

\newtheorem{problem}[equation]{Problem}
\newtheorem{example}[equation]{Example}
\newtheorem{exercise}[equation]{Exercise}
\newtheorem*{answer}{Answer}
\newtheorem*{solution}{Solution}
\newtheorem{remark}[equation]{Remark}
\newtheorem{terminology}[equation]{Terminology}

\newtheorem{notation}[equation]{Notation}
\newtheorem{noterm}[equation]{Notation and Terminology}

\newcommand\define[1]{\emph{\textbf{#1}}}

\numberwithin{equation}{section}

\newcommand{\be}{\begin{equation}}
\newcommand{\ee}{\end{equation}}
\newcommand{\bx}{\begin{example}}
\newcommand{\ex}{\end{example}}
\newcommand{\bex}{\begin{exercise}}
\newcommand{\eex}{\end{exercise}}
\newcommand{\ban}{\begin{answer}}
\newcommand{\ean}{\end{answer}}
\newcommand{\bt}{\begin{theorem}}
\newcommand{\et}{\end{theorem}}
\newcommand{\bc}{\begin{corollary}}
\newcommand{\ec}{\end{corollary}}
\newcommand{\blem}{\begin{lemma}}
\newcommand{\elem}{\end{lemma}}
\newcommand{\bp}{\begin{problem}}
\newcommand{\ep}{\end{problem}}
\newcommand{\bn}{\begin{proposition}}
\newcommand{\en}{\end{proposition}}
\newcommand{\bd}{\begin{definition}}
\newcommand{\ed}{\end{definition}}
\newcommand{\bq}{\begin{question}}
\newcommand{\eq}{\end{question}}
\newcommand{\bprf}{\begin{proof}}
\newcommand{\eprf}{\end{proof}}
\newcommand{\br}{\begin{remark}}
\newcommand{\er}{\end{remark}}
\newcommand{\bs}{\begin{solution}}
\newcommand{\es}{\end{solution}}
\newcommand{\bnt}{\begin{noterm}}
\newcommand{\ent}{\end{noterm}}
\newcommand{\bnot}{\begin{notation}}
\newcommand{\enot}{\end{notation}}
\newcommand{\bterm}{\begin{terminology}}
\newcommand{\eterm}{\end{terminology}}

\newcommand{\<}{\langle}
\renewcommand{\>}{\rangle}

\newcommand{\id}{\mathrm{id}}

\newcommand{\Tr}{{\rm Tr} }

\def\C{{{\mathbb C}}}

\DeclareMathAlphabet{\mathpzc}{OT1}{pzc}{m}{it} 
 \DeclareFontFamily{OT1}{pzc}{}
 \DeclareFontShape{OT1}{pzc}{m}{it}{ <-> s*[1.2] pzcmi7t }{}
 \DeclareMathAlphabet{\mathpzc}{OT1}{pzc}{m}{it}
 \newcommand{\Alg}[1]{\mathpzc{#1}}

\def\A{\Alg{A}}
\def\B{\Alg{B}}

\def\E{\mathcal{E}}
\def\H{\mathcal{H}}

\def\O{\mathscr{O}}

\def\J{\mathscr{J}}
\def\C{\mathbb{C}}

\title{On Dirac-type correlations}
\author{James Fullwood}
\affiliation{School of Mathematics and Statistics, Hainan University, Haikou, Hainan, 570228, China}
\email{fullwood@hainanu.edu.cn}
\author{Boyu Yang}
\affiliation{School of Mathematics and Statistics, Hainan University, Haikou, Hainan, 570228, China}

\newcommand{\Addresses}{{
}}

\begin{document}
\emergencystretch 2em

\maketitle


\vspace{-7mm}
\tableofcontents

\begin{abstract}
Quantum correlations often defy an explanation in terms of fundamental notions of classical physics, such as causality, locality, and realism. While the mathematical theory underpinning quantum correlations between spacelike separated systems has been well-established since the 1930s, the mathematical theory for correlations between non-spacelike separated systems is much less developed. In this work, we develop the theory of what we refer to as \emph{local-density operators}, which we view as joint states for possibly non-spacelike separated quantum systems. Local-density operators are unit trace operators whose marginals are genuine density operators, which we show not only subsumes the notion of density operator, but also several extensions of the notion of density operator into the spatiotemporal domain, such as pseudo-density operators and quantum states over time. More importantly, we prove a result which establishes a one-to-one correspondence between local-density operators and what we refer to as \emph{Dirac measures}, which are complex-valued measures on the space of separable projectors associated with two quantum systems. In the case that one of the systems is the trivial quantum system with Hilbert space $\C$, our result recovers the fundamental result known as Gleason's Theorem, which implies that the Born rule from quantum theory is the only way in which one may assign probabilities to the outcomes of measurements performed on quantum systems in a non-contextual manner. As such, our results establish a direct generalization of Gleason's Theorem to measurements performed on possibly non-spacelike separated systems, thus extending the mathematical theory of quantum correlations across space to quantum correlations across space \emph{and} time.
\end{abstract}

	\maketitle    

\section{Introduction}

In quantum theory, the correlation between observables $\O_A$ and $\O_B$ measured on quantum systems $A$ and $B$ is defined to be the expectation value of the \emph{product} of the measurements of $\O_A$ and $\O_B$. If $A$ and $B$ are spacelike separated and the composite system $AB$ is in a state represented by a density operator $\rho_{AB}$ prior to measurement, the associated correlation $\mathscr{C}(\O_A,\O_B)$ is then given by 
\[
\mathscr{C}(\O_A,\O_B)=\sum_{i,j}\lambda_i\nu_j\bold{P}(i,j)\, ,
\]
where $\O_A=\sum_i \lambda_iP_i$ and $\O_B=\sum_j \nu_jQ_j$ are spectral decompositions of the observables $\O_A$ and $\O_B$, and $\bold{P}(i,j)=\Tr[\rho_{AB}(P_i\otimes Q_j)]$ is the probability of the joint outcome $P_i$ and $Q_j$ as given by the Born rule. By the linearity of trace, it follows that 
\be \label{EXTVZ87}
\mathscr{C}(\O_A,\O_B)=\Tr[\rho_{AB}(\O_A\otimes \O_B)]\, ,
\ee
so that the correlation $\mathscr{C}(\O_A,\O_B)$ is simply the usual expectation value of the product observable $\O_A\otimes \O_B$. 

On the other hand, if $\O_A$ and $\O_B$ are instead measured in sequence on a \emph{single} quantum system---so that $A$ and $B$ now denote the same quantum system at two times $t_A$ and $t_B>t_A$---then the issue of giving a theoretical definition of the expectation value of the product of the measurements of $\O_A$ followed by $\O_B$ becomes problematic. In particular, if $\{P_i\}$ and $\{Q_j\}$ are projective measurements instantiating a determination of the values of the observables $\O_A$ and $\O_B$, respectively, then according to the L\"{u}ders-von~Neumann projection postulate~\cite{Lu06,vN18}, the joint probability $\bold{P}(i,j)$ of outcome $P_i$ \emph{followed} by $Q_j$ is given by
\be \label{PTMXS71}
\bold{P}(i,j)=\Tr[P_i\rho_AP_iQ_j]\, ,
\ee
where $\rho_A$ is a density operator representing the initial state of the system prior to measurement. In Ref.~\cite{FullwoodWu_2025}, it was shown that if $\rho_A$ is not the maximally mixed state---i.e., when there exists a projector $P_i$ which does not commute with $\rho_A$---then there \emph{does not} exist an operator $\varrho_{AB}$ such that the probabilities $\bold{P}(i,j)$ may be computed in terms of $\Tr[\varrho_{AB}(P_i\otimes Q_j)]$ as $\{P_i\}$ and $\{Q_j\}$ vary across all projective measurements. As such, it follows that the correlation between $A$ and $B$ associated with expectation value $\sum_{i,j}\lambda_i\nu_j\bold{P}(i,j)$ depends on the projective measurements $\{P_i\}$ and $\{Q_j\}$ instantiating a determination of the values of the observables $\O_A$ and $\O_B$, which are non-unique whenever $\O_A$ or $\O_B$ have degenerate eigenvalues. Such a dependence on the projective measurements instantiating a determination of the values of the observables $\O_A$ and $\O_B$ is in stark contrast to the case when $A$ and $B$ are spacelike separated, since in such a case the correlation $\mathscr{C}(\O_A,\O_B)$ is well-defined irrespective of whether or not the observables $\O_A$ and $\O_B$ have degenerate eigenvalues, and is thus independent of the way in which $\O_A$ and $\O_B$ are measured.

In 1945, Dirac defined a correlation function for timelike separated observables in terms of complex-valued probabilities $\bold{Q}(i,j)$ taking the place of the probabilities $\bold{P}(i,j)$ as given by \eqref{PTMXS71}~\cite{Dirac_1945}. As such, the Dirac correlation function $\mathscr{D}(\O_A,\O_B)$ associated with timelike separated observables $\O_A=\sum_i \lambda_i P_i$ and $\O_B=\sum_j \nu_j Q_j$ is the complex expectation value given by
\[
\mathscr{D}(\O_A,\O_B)=\sum_{i,j}\lambda_i\nu_j \bold{Q}(i,j)\, ,
\]
and it is immediate from the definition of $\bold{Q}(i,j)$ that such an expectation value is independent of the projective measurements $\{P_i\}$ and $\{Q_j\}$ instantiating a determination of the values of the observables $\O_A$ and $\O_B$, in accordance with the spacelike correlation $\mathscr{C}(\O_A,\O_B)$. The Dirac correlation function $\mathscr{D}(\O_A,\O_B)$ has since become fundamental to the study of quantum field theory and quantum thermodynamics, where it is often referred to as the \emph{two-point correlator}. What is less known however, is that the Dirac correlation function also played a crucial role in motivating Feynman's formulation of the path integral~\cite{Feynman_1948}.

As the Dirac correlation function is independent of the projective measurements which implement the determination of the values of observables $\O_A$ and $\O_B$, it is straightforward to show that there exists a unique bipartite operator $\varrho_{AB}$ such that for all observables $\O_A$ and $\O_B$,
\be \label{DRXSCS17}
\mathscr{D}(\O_A,\O_B)=\Tr[\varrho_{AB}(\O_A\otimes \O_B)]\, ,
\ee
so that the Dirac correlation function may be given the same form as that of \eqref{EXTVZ87} for the spacelike correlation function $\mathscr{C}(\O_A,\O_B)$. As such, the operator $\varrho_{AB}$ has an operational meaning which is analogous to that of a density operator $\rho_{AB}$, and thus may be viewed as a generalized quantum state. For these reasons, the operator $\varrho_{AB}$ has been referred to as a `quantum state over time' in Refs.~\cite{HHPBS17,FuPa22,FuPa22a,LiNg23,LieFu_2025}, and as a `spacetime density matrix' in Refs.~\cite{Diaz_25,Diaz_24,Milekhin2025,Guo_2025}.   Moreover, the fact that the Dirac correlation function may be given by \eqref{DRXSCS17} stems from the fact that the Dirac probabilities $\bold{Q}(i,j)$ may also be given by 
\be \label{BXR757}
\bold{Q}(i,j)=\Tr[\varrho_{AB}(P_i\otimes Q_j)]\, ,
\ee
which may be viewed as a generalization of the Born rule to non-spacelike separated measurements. 

Contrary to density operators, the operator $\varrho_{AB}$ encoding the Dirac correlation function is not Hermitian in general, let alone positive semi-definite. However, in accordance with bipartite density operators, the operator $\varrho_{AB}$ satisfies the marginal conditions 
\be \label{MCXND91}
\Tr_B[\varrho_{AB}]=\rho_A \quad \text{and} \quad \Tr_A[\varrho_{AB}]=\rho_B\, ,
\ee
where $\rho_A$ and $\rho_B$ are reduced density operators representing quantum states of the local systems $A$ and $B$. Now just as an arbitrary bipartite density operator $\rho_{AB}$ serves as a possible quantum state for spacelike separated systems, here we take the viewpoint that an arbitrary operator $\varrho_{AB}$ satisfying the marginal conditions \eqref{MCXND91} may serve as a quantum state for possibly non-spacelike separated systems.

In this work, we then define a \emph{local-density operator} to be any operator $\varrho_{AB}$ satisfying the marginal conditions \eqref{MCXND91}, and we develop a general theory of local-density operators and their associated correlation functions as defined by the RHS of equation \eqref{DRXSCS17}.  In particular, we show that the notion of local-density operator subsumes not only the notion of density operator, but also several extensions of the notion of density operator into the temporal domain, such as that of pseudo-density operators~\cite{FJV15}, causal states~\cite{LeSp13}, and quantum states over time~\cite{HHPBS17}. As such, local-density operators admit various operational meanings, such as encoding the correlations between prepare-evolve-measure scenarios~\cite{FuPa22a}, sequential measurement scenarios~\cite{FuPa24}, post-selection~\cite{PaTK23}, and the interference terms of interferometry~\cite{Lie_2025}. We also prove that every local-density operator uniquely defines a complex-valued measure which we refer to as a \emph{Dirac measure} on the space of separable projectors 
\be \label{SPLPRJXT769}
\left.\big\{P\otimes Q\,\right|\,\text{$P$ and $Q$ are projectors on $A$ and $B$, respectively}\big\}\, ,
\ee
and we show that every such Dirac measure is induced from a unique local-density operator via a trace formula as in \eqref{BXR757}. If we take one of the systems $A$ or $B$ to be the trivial quantum system associated with the Hilbert space $\C$, we recover the fundamental result of Gleason~\cite{Gleason_57}, which states that every generalized probability measure on the space of projectors corresponding to a fixed quantum system is induced by the Born rule associated with a unique density operator. As such, our result may then be viewed as a generalization of Gleason's Theorem to local measurements being performed on possibly non-spacelike separated quantum systems $A$ and $B$. Although an extension of our result to local measurements being performed on any finite number of systems is straightforward, we focus on the case of two local measurements for conceptual clarity.

While it is usually assumed that the observables appearing in the Dirac correlation function \eqref{DRXSCS17} are ordered in time, for the correlation function associated with a general local-density operator $\varrho_{AB}$ there is no need to assume any fixed spacetime labels associated with the systems $A$ and $B$. In particular, it has recently been shown that there exists a class of bipartite density operators whose associated correlations may be realized either by spacelike separated or timelike separated systems~\cite{song23}, thus even in the case of positive local-density operators the assignment of spacetime labels may be ambiguous, or observer-dependent. As such, we may view a general local-density operator $\varrho_{AB}$ as encoding abstract correlations between quantum systems $A$ and $B$ in a way which does not make any a priori assumption about the manner in which $A$ and $B$ may be situated in spacetime, thus providing a foundation for a background independent formulation of quantum theory which treats space and time on equal footing. 

The rest of this paper is organized as follows. In Section~\ref{S2}, we formally introduce the notion of a Dirac measure on the space of separable projectors \eqref{SPLPRJXT769}, and we show that every Dirac measure induces a unique bilinear correlation function between observables on possibly non-spacelike separated systems $A$ and $B$. In Section~\ref{S3} we prove a generalization of Gleason's Theorem for Dirac measures. In particular, we prove that there is a one-to-one correspondence between Dirac measures and local-density operators, and that such local-density operators provide a unique linear representation of the correlation functions associated with Dirac measures. In Section~\ref{S4} we recall various examples of Dirac measures from the literature, and we give explicit formulas for their associated local-density operators. We go over in detail the operational interpretation of such Dirac measures and their associated local-density operators, and we also provide an example of a family of local-density operators whose associated correlations are not yet well-understood from an operational standpoint. In Section~\ref{S5}, we derive a Bayes' rule associated with every Dirac measure by making use of a quantum Bayes' rule which was first introduced in Ref.~\cite{FuPa22a}, which is an operator equation from which one may derive the classical Bayes' rule.  

\section{Dirac measures and their associated correlation functions} \label{S2}

Throughout this work, we consider two quantum systems $A$ and $B$ with finite-dimensional Hilbert spaces $\H_A$ and $\H_B$, and we consider local measurements being performed independently on systems $A$ and $B$. The algebra of linear operators on $\H_A$ will be denoted by $\A$, and the algebra of linear operators on $\H_B$ will be denoted by $\B$. The associated composite system will be denoted by $AB$, and given $X\in\{A,B,AB\}$, the set of density operators on $\H_X$ will be denoted by $\mathfrak{D}(X)$, the set of projectors (or idempotents) on $\H_X$ will be denoted by $\bold{Proj}(X)$, the space of Hermitian operators on $\H_X$ will be denoted by $\bold{Obs}(X)$, and the identity operator on $\H_X$ will be denoted by $\mathds{1}_X$. A collection $\{P_i\}\subset \bold{Proj}(X)$ of mutually orthogonal projectors which sum to the identity operator on $\H_X$ will be referred to as a \define{projective measurement}, or a \define{projection-valued measure} (PVM) on $X$. Given an observable $\O\in \bold{Obs}(X)$, a decomposition $\O=\sum_i \lambda_i P_i$ will be referred to as a \define{spectral decomposition} of $\O$ if and only if $\{P_i\}$ is a projective measurement. We then let $\bold{Proj}(A,B)\subset \bold{Proj}(AB)$ denote the subset given by
\[
\bold{Proj}(A,B)=\left.\big\{P\otimes Q\,\,\right|\,\,\text{$P\in \bold{Proj}(A)$ and $Q\in \bold{Proj}(B)$} \big\}\, ,
\]
so that $\bold{Proj}(A,B)$ corresponds to the fundamental set of conjunctive propositions associated with local projective measurements being performed independently on systems $A$ and $B$~\cite{Zeilinger_1999}. 

We now define what we refer to as a \emph{Dirac measure} on $\bold{Proj}(A,B)$, which generalizes a notion of complex-valued probability as first introduced by Kirkwood in 1933~\cite{Kirkwood_1933} and later by Dirac in 1945~\cite{Dirac_1945}. While the notion of a Dirac measure is complex-valued, we note that it doesn't rule out the possibility of real-valued or even positive-valued measures. In particular, the notion of Dirac measure also encompasses the notion of joint probabilities associated with spacelike separated systems. 

\bd
A function $\mu:\bold{Proj}(A,B)\to \C$ is said to be a \define{Dirac measure} if and only if the following conditions are satisfied:
\begin{itemize}
\item
(Normalization) $\mu(\mathds{1}_A\otimes \mathds{1}_B)=1$.
\item
(Boundedness) $\sup\{|\mu(P\otimes Q)|\,|\, P\otimes Q\in \bold{Proj}(A,B)\}<\infty$.
\item
(Local Positivity) $\mu(P\otimes \mathds{1}_B)\geq 0$ and $\mu(\mathds{1}_A\otimes Q)\geq 0$ for all $P\in \bold{Proj}(A)$ and $Q\in \bold{Proj}(B)$.
\item
(Local Additivity) For every mutually orthogonal collection $\{P_i\}\subset \bold{Proj}(A)$ and $\{Q_j\}\subset \bold{Proj}(B)$,
\[
\mu\Big(\Big(\sum_i P_i\Big)\otimes Q\Big)=\sum_i\mu(P_i\otimes Q) \qquad \forall Q\in \bold{Proj}(B)\, , 
\] 
and
\[
\mu\Big(P\otimes \Big(\sum_j Q_j\Big)\Big)=\sum_j\mu(P\otimes Q_j) \qquad \forall P\in \bold{Proj}(A)\, . 
\] 
\end{itemize}
The set of all such Dirac measures will be denoted by $\mathfrak{Dirac}(AB)$. If a Dirac measure $\mu$ is such that $\mu(P\otimes Q)\geq 0$ for all $P\otimes Q\in \bold{Proj}(A,B)$, then $\mu$ will be referred to as a \define{positive Dirac measure}. The set of all such positive Dirac measures will be denoted by $\mathfrak{Dirac}^+(AB)$.
\ed 

In Section~\ref{S4} we provide several examples of Dirac measures which have appeared in the literature, though they were not referred to as such. The key properties of a Dirac measure are boundedness and local additivity, which imply that a Dirac measure uniquely determines a well-defined, complex-valued correlation function associated with measurements of local observables performed on systems $A$ and $B$, as we now show. 

\bt \label{TMXS17}
Suppose $\dim(\mathcal{H}_A)\neq 2 \neq \dim(\mathcal{H}_B)$, and let $\mu:\bold{Proj}(A,B)\to \C$ be a Dirac measure. Then there exists a unique bilinear functional $\mathscr{D}_{\mu}:\bold{Obs}(A)\times \bold{Obs}(B)\to \C$ such that $\mathscr{D}_{\mu}(P,Q)=\mu(P\otimes Q)$ for all $P\otimes Q\in \bold{Proj}(A,B)$. Moreover, the bilinear functional $\mathscr{D}_{\mu}:\bold{Obs}(A)\times \bold{Obs}(B)\to \C$ associated with a Dirac measure $\mu:\bold{Proj}(A,B)\to \C$ is given by
\be \label{CFXN89}
\mathscr{D}_{\mu}(\O_A,\O_B)=\sum_{i,j}\lambda_i \nu_j\, \mu(P_i\otimes Q_j)\, ,
\ee
where $\O_A=\sum_i \lambda_i P_i$ and $\O_B=\sum_j \nu_j Q_j$ are spectral decompositions of $\O_A$ and $\O_B$.
\et

\br
If an observable $\O\in \bold{Obs}(X)$ with $X\in \{A,B\}$ has a degenerate eigenvalue, then a spectral decomposition of $\O$ is highly non-unique due to the continuum of ways in which one may choose an orthonormal basis of the eigenspace associated with a degenerate eigenvalue. As such, it is not clear a priori if a function $\mathscr{D}_{\mu}$ as given by \eqref{CFXN89} is well-defined, as it is defined in terms of spectral decompositions of observables $\O_A$ and $\O_B$. Therefore, the non-trivial part of Theorem~\ref{TMXS17} is to show there exists a bilinear functional $\mathscr{D}_{\mu}:\bold{Obs}(A)\times \bold{Obs}(B)\to \C$ such that $\mathscr{D}_{\mu}(P,Q)=\mu(P\otimes Q)$ for all $P\otimes Q\in \bold{Proj}(A,B)$, which then implies $\mathscr{D}_{\mu}$ may be given by the formula \eqref{CFXN89}.
\er

Our proof of Theorem~\ref{TMXS17} relies on the following lemma, which follows from the \emph{Mackey--Gleason theorem} as proved by Bunce and Wright~\cite{Bunce1992}.
\begin{lemma}
\label{complex-gleasonthm}
Let $\mathcal{H}$ be a finite-dimensional complex Hilbert space with $\dim(\mathcal{H}) \neq 2$, and suppose $\mu : \bold{Proj}(\mathcal{H}) \to \mathbb{C}$ is a bounded function---i.e., $\sup\{|\mu(P)|\,|\, P\in \bold{Proj}(\mathcal{H})\}<\infty$---such that for every mutually orthogonal collection $\{P_i\}\subset \bold{Proj}(\mathcal{H})$, 
\be \label{ORTXADXT71}
\mu\Big(\sum_i P_i\Big)=\sum_i\mu(P_i)\, .
\ee
Then there exists a unique operator $\tau$ on $\mathcal{H}$ such that $\mu(P) = \Tr[\tau P]$, for all $P \in \bold{Proj}(\mathcal{H})$.
\end{lemma}

\bprf[Proof of Theorem~\ref{TMXS17}] 
Since $\mu$ is a Dirac measure, it is bounded, so there exists $M>0$ such that $|\mu(P\otimes Q)|\leq M$ for all $P\otimes Q\in \bold{Proj}(A,B)$. For every $P\in \bold{Proj}(A)$ and $Q\in \bold{Proj}(B)$, let $\mu_P:\bold{Proj}(B)\to \C$ and $\mu^Q:\bold{Proj}(A)\to \C$ be the maps given by
\[
\mu_P(Q)=\mu(P\otimes Q) \qquad \forall Q\in \bold{Proj}(B)\, ,
\]
and
\[
\mu^Q(P)=\mu(P\otimes Q) \qquad \forall P\in \bold{Proj}(A)\, .
\]
Clearly $|\mu_P(Q)|\leq M$ and $|\mu^Q(P)|\leq M$ for all $P\in \bold{Proj}(A)$ and $Q\in \bold{Proj}(B)$, so $\mu_P$ and $\mu^Q$ are bounded. Since local additivity of $\mu$ implies that $\mu_P$ and $\mu^Q$ satisfy the additivity condition \eqref{ORTXADXT71} of Lemma~\ref{complex-gleasonthm}, it follows that there exist unique operators $\tau^Q\in \A$ and $\tau_P\in \B$ such that $\mu_P(Q)=\Tr[\tau_P Q]$ for all $Q\in \bold{Proj}(B)$ and $\mu^Q(P)=\Tr[\tau^QP]$ for all $P\in \bold{Proj}(A)$. Now given $\O_B\in \bold{Obs}(B)$, we let $\mu_{\O_B}:\bold{Proj}(A)\to \C$ be the map given by
\[
\mu_{\O_B}(P)=\Tr[\tau_P \O_B] \qquad \forall P\in \bold{Proj}(A)\, .
\]
Fixing a spectral decomposition $\O_B=\sum_{j}\nu_j Q_j$, we have for every $P\in \bold{Proj}(A)$ that
\[
|\mu_{\O_B}(P)|=\Big|\sum_j \nu_j \Tr[\tau_P Q_j]\Big|=\Big|\sum_j \nu_j \mu(P\otimes Q_j)\Big|\leq M\sum_j |\nu_j|\, ,
\]
so $\mu_{\O_B}$ is bounded on $\bold{Proj}(A)$. Moreover, if $\{P_i\}\subset \bold{Proj}(A)$ is a mutually orthogonal collection and $P=\sum_i P_i$, we have
\begin{align*}
\mu_{\mathscr{O}_B}\Big(\sum_i P_i\Big)&=\mu_{\mathscr{O}_B}(P)=\Tr[\tau_{P}\O_B]=\Tr\Big[\tau_P \sum_j\nu_j Q_j\Big] \\
&=\sum_j \nu_j\Tr[\tau_P Q_j]=\sum_j \nu_j \mu_P(Q_j)=\sum_j \nu_j \mu(P\otimes Q_j) \\
&=\sum_j \nu_j \mu\Big(\Big(\sum_i P_i\Big) \otimes Q_j\Big)=\sum_j \nu_j \sum_i\mu(P_i \otimes Q_j) \\
&=\sum_j \nu_j \sum_i\Tr[\tau_{P_i} Q_j]=\sum_i\Tr\Big[\tau_{P_i} \sum_j \nu_j Q_j\Big] \\
&=\sum_i\Tr[\tau_{P_i} \O_B]=\sum_i\mu_{\O_B}(P_i)\, ,
\end{align*}
thus $\mu_{\O_B}$ satisfies both the boundedness and additivity conditions of Lemma~\ref{complex-gleasonthm}. It then follows that there exists a unique operator $\varphi_{\O_B}\in \A$ such that
\[
\Tr[\varphi_{\O_B}P]=\mu_{\O_B}(P)=\Tr[\tau_P \O_B] \qquad \forall P\in \bold{Proj}(A)\, .
\]
Now let $\mathscr{D}_{\mu}:\bold{Obs}(A)\times \bold{Obs}(B)\to \C$ be the map given by
\[
\mathscr{D}_{\mu}(\O_A,\O_B)=\Tr[\varphi_{\O_B}\O_A] \qquad \forall \O_A\in \bold{Obs}(A)\, ,\, \O_B\in \bold{Obs}(B)\, .
\]
Then for every $P\in \bold{Proj}(A)$ and $Q\in \bold{Proj}(B)$ we have
\[
\mathscr{D}_{\mu}(P,Q)=\Tr[\varphi_Q P]=\Tr[\tau_P Q]=\mu(P\otimes Q)\, ,
\]
thus $\mathscr{D}_{\mu}(P,Q)=\mu(P\otimes Q)$ for all $P\otimes Q\in \bold{Proj}(A,B)$. Moreover, $\mathscr{D}_{\mu}$ is certainly linear with respect to $\bold{Obs}(A)$ by the linearity of trace, so to show $\mathscr{D}_{\mu}$ is bilinear it suffices to show that $\mathscr{D}_{\mu}$ is linear with respect to $\bold{Obs}(B)$. For this, let $\Phi:\bold{Obs}(B)\to \A$ be the map given by 
\[
\Phi(\O_B)=\varphi_{\O_B} \qquad \forall \O_B\in \bold{Obs}(B)\, ,
\]
so that $\mathscr{D}_{\mu}$ is linear with respect to $\bold{Obs}(B)$ if and only if $\Phi$ is linear. Given a real-linear combination $\nu_1\O_B^{(1)}+\nu_2\O_B^{(2)}$ of observables in $\bold{Obs}(B)$, we have for every $P\in \bold{Proj}(A)$ that
\begin{align*}
\Tr\left[\Phi\Big(\nu_1\O_{B}^{(1)}+\nu_2\O_{B}^{(2)}\Big)P\right]&=\Tr\left[\varphi_{\nu_1\O _{B}^{(1)}+\nu_2\O_{B}^{(2)}}P\right] \\
&=\Tr\left[\tau_P\big(\nu_1\O_{B}^{(1)}+\nu_2\O_B^{(2)}\big)\right] \\
&=\nu_1\Tr\left[\tau_P\O_B^{(1)}\right]+\nu_2\Tr\left[\tau_P\O_B^{(2)}\right] \\
&=\nu_1\Tr\left[\varphi_{\O _{B}^{(1)}}P\right]+\nu_2\Tr\left[\varphi_{\O_{B}^{(2)}}P\right] \\
&=\Tr\left[\Big(\nu_1\Phi\big(\O_{B}^{(1)}\big)+\nu_2\Phi\big(\O_{B}^{(2)}\big)\Big)P\right]\, .
\end{align*}
Since $\bold{Proj}(A)$ linearly spans $\A$ and the Hilbert--Schmidt inner product is nondegenerate, it follows that $\Phi\big(\nu_1\O_B^{(1)}+\nu_2\O_B^{(2)}\big)=\nu_1\Phi\big(\O_B^{(1)}\big)+\nu_2\Phi\big(\O_B^{(2)}\big)$, thus $\mathscr{D}_{\mu}$ is indeed linear with respect to $\bold{Obs}(B)$, and hence bilinear. Moreover, since $\mathscr{D}_{\mu}$ is a bilinear functional such that $\mathscr{D}_{\mu}(P,Q)=\mu(P\otimes Q)$ for all $P\otimes Q\in \bold{Proj}(A,B)$, it follows that $\mathscr{D}_{\mu}$ is necessarily given by \eqref{CFXN89}, and is thus unique.
\eprf

\bd
Given a Dirac measure $\mu:\bold{Proj}(A,B)\to \C$, the bilinear functional $\mathscr{D}_{\mu}:\bold{Obs}(A)\times \bold{Obs}(B)\to \C$ as given by \eqref{CFXN89} will be referred to as the \define{correlation function} associated with the Dirac measure $\mu$.
\ed

\section{A generalization of Gleason's Theorem \label{S3}} 

In 1957, Gleason proved what is known today as \emph{Gleason's Theorem}~\cite{Gleason_57}, which is a fundamental result showing that the Born rule of quantum theory is the only way in which probabilities may be assigned to measurement outcomes on a quantum system in a non-contextual manner. For the precise statement, let $X$ be a quantum system with $\text{dim}(\H_X)\neq 2$, and suppose that a function $\mu:\bold{Proj}(X)\to [0,1]$ satisfies the following properties:
\begin{itemize}
\item
(Normalization) $\mu(\mathds{1}_X)=1$.
\item
(Orthogonal Additivity) If $\{P_i\}\subset \bold{Proj}(X)$ is a mutually orthogonal collection, so that $P_iP_j=\delta_{ij}P_i$, then
\[
\mu\Big(\sum_iP_i\Big)=\sum_i\mu(P_i)\, .
\]
\end{itemize}
Gleason's Theorem then states that there exists a unique density operator $\rho\in \mathfrak{D}(X)$ such that $\mu(P)=\Tr[\rho P]$ for all $P\in \bold{Proj}(X)$. Note that if $X=AB$ is a composite system with $B$ the trivial quantum system with Hilbert space $\C$, then we may identify $\bold{Proj}(A,B)$ with $\bold{Proj}(X)$, in which case such a function $\mu:\bold{Proj}(X)\to [0,1]$ is in fact a Dirac measure. As such, a Dirac measure may be viewed as a generalization of the probability measures $\mu$ characterized by Gleason's Theorem. In this section, we then prove the corresponding generalization of Gleason's Theorem for Dirac measures. The analog of a density operator in such a case is then given by the following:

\bd
An operator $\varrho_{AB}\in \A\otimes \B$ will be referred to as a \define{local-density operator} if and only if $\Tr_B[\varrho_{AB}]\in \mathfrak{D}(A)$ and $\Tr_A[\varrho_{AB}]\in \mathfrak{D}(B)$. The set of all such local-density operators in $\A\otimes \B$ will be denoted by $\mathfrak{Loc}(AB)$ (note that $\mathfrak{D}(AB)\subset \mathfrak{Loc}(AB)$).
\ed

\bt \label{MTX1}
Let $\mu:\bold{Proj}(A,B)\to \C$ be a function, and suppose $\emph{dim}(\mathcal{H}_A)\neq 2 \neq \emph{dim}(\mathcal{H}_B)$. Then $\mu$ is a Dirac measure if and only if there exists a unique local-density operator $\varrho_{AB}\in \mathfrak{Loc}(AB)$ such that
\be \label{GLEASX87}
\mu(P\otimes Q)=\Tr[\varrho_{AB}(P\otimes Q)] \qquad \forall P\otimes Q\in \bold{Proj}(A,B)\, .
\ee
Moreover, the mapping $f:\mathfrak{Dirac}(AB)\to \mathfrak{Loc}(AB)$ given by $f(\mu)=\varrho_{AB}$ is a bijection.
\et

\bprf
$(\implies)$ Suppose $\mu$ is a Dirac measure, and let $\mathscr{D}_{\mu}:\bold{Obs}(A)\times \bold{Obs}(B)\to \C$ be the correlation function associated with $\mu$ as given by \eqref{CFXN89}. Since $\mathscr{D}_{\mu}$ is bilinear, and since every linear operator $X\in \A$ is of the form $X=H_1+iH_2$ with $H_1$ and $H_2$ Hermitian (and similarly for any operator $Y\in \B$), it follows that $\mathscr{D}_{\mu}$ may be uniquely extended to a bilinear functional $\mathscr{D}_{\mu}':\A\times \B\to \C$. It then follows from the Riesz representation theorem that there exists a unique operator $\varrho_{AB}$ such that 
\[
\mathscr{D}_{\mu}'(X,Y)=\Tr[\varrho_{AB}(X\otimes Y)] \qquad \forall X\in \A\, ,\, Y\in \B\, .
\]
We now show that $\varrho_{AB}$ is in fact a local-density operator. Indeed, since
\[
\Tr[\varrho_{AB}]=\Tr[\varrho_{AB}(\mathds{1}_A\otimes \mathds{1}_B)]=\mathscr{D}_{\mu}'(\mathds{1}_A,\mathds{1}_B)=\mathscr{D}_{\mu}(\mathds{1}_A,\mathds{1}_B)=\mu(\mathds{1}_A\otimes \mathds{1}_B)=1\, ,
\]
it follows that $\rho_A=\Tr_B[\varrho_{AB}]$ and $\rho_B=\Tr_A[\varrho_{AB}]$ are both of unit trace. Moreover, for every $P\in \bold{Proj}(A)$ we have
\[
\Tr[\rho_A P]=\Tr[\varrho_{AB}(P\otimes \mathds{1}_B)]=\mu(P\otimes \mathds{1}_B)\geq 0\, ,
\]
from which it follows that $\rho_A$ is positive semi-definite, and hence $\rho_A\in \mathfrak{D}(A)$. Similarly, we have $\Tr[\rho_B Q]\geq 0$ for all $Q\in \bold{Proj}(B)$, thus $\rho_B\in \mathfrak{D}(B)$, and hence $\varrho_{AB}$ is a local-density operator, as desired.

$(\impliedby)$ Suppose there exists a local-density operator $\varrho_{AB}\in \mathfrak{Loc}(AB)$ such that \eqref{GLEASX87} holds for all $P\otimes Q\in \bold{Proj}(A,B)$. Then clearly $\mu$ is normalized, locally positive, and locally additive. Moreover, by the Cauchy--Schwarz inequality for the Hilbert--Schmidt inner product, for all $P\otimes Q\in \bold{Proj}(A,B)$ we have
\[
|\mu(P\otimes Q)|=|\Tr[\varrho_{AB}(P\otimes Q)]|\leq \sqrt{\Tr[\varrho_{AB}^\dagger\varrho_{AB}]\,\Tr[P\otimes Q]}\leq \sqrt{\dim(\mathcal{H}_A)\dim(\mathcal{H}_B)\,\Tr[\varrho_{AB}^\dagger\varrho_{AB}]}\, ,
\]
so $\mu$ is bounded. It then follows that $\mu$ is a Dirac measure, as desired.

We now show that the map $f:\mathfrak{Dirac}(AB)\to \mathfrak{Loc}(AB)$ given by $f(\mu)=\varrho_{AB}$ is in fact a bijection. Certainly $f$ is surjective by the implication $(\impliedby)$, thus it suffices to show $f$ is injective. For this, suppose $\varrho_{AB},\varrho_{AB}'\in \mathfrak{Loc}(AB)$
induce the same Dirac measure $\mu$, so that
\[
\Tr[(\varrho_{AB}-\varrho_{AB}')(P\otimes Q)]=0
\qquad
\forall P\otimes Q\in \bold{Proj}(A,B)\, .
\]
Since elements of the form $P\otimes Q\in \bold{Proj}(A,B)$ linearly span
$\bold{Obs}(A)\otimes\bold{Obs}(B)$, and since the Hilbert--Schmidt inner product is nondegenerate, it follows that $\varrho_{AB}=\varrho_{AB}'$. Therefore, $f$ is injective, and hence bijective, as desired.
\eprf

In light of Theorem~\ref{MTX1}, the notion of local-density operator $\varrho_{AB}$ generalizes the notion of a joint density operator $\rho_{AB}$ to composite systems $AB$ where $A$ and $B$ are possibly non-spacelike separated. In particular, local-density operators encode general Dirac measures in precisely the same way that bipartite density operators encode positive Dirac measures associated with spacelike separated systems. As such, a local-density operator $\varrho_{AB}$ will be referred to as a \define{state} of the composite system $AB$. While a local-density operator is necessarily of unit trace, a general local-density operator is not necessarily positive semi-definite, or even Hermitian. However, if we restrict our attention to real-valued Dirac measures, then Theorem~\ref{MTX1} establishes a bijective correspondence between real-valued Dirac measures and Hermitian local-density operators.

We note that although the space $\bold{Proj}(A,B)$ of separable projectors is much smaller than the entire space $\bold{Proj}(AB)$ of projectors on the composite system $AB$, the four simple conditions of normalization, boundedness, local positivity, and local additivity nevertheless are enough to single out a unique local-density operator associated with a Dirac measure. This is in contrast to Gleason's Theorem (or Lemma~\ref{complex-gleasonthm}), which requires orthogonal additivity on the entire space $\bold{Proj}(AB)$ to establish a one-to-one correspondence between probability measures and bipartite density operators. In any case, setting $B$ to be the trivial quantum system with Hilbert space $\C$ recovers Gleason's Theorem for system $A$, thus Theorem~\ref{MTX1} may be viewed as a generalization of Gleason's Theorem to local measurements performed on possibly non-spacelike separated systems. 

We also note that a result similar to Theorem~\ref{MTX1} was proved by Wallach, which is referred to as an ``unentangled Gleason's theorem" in Ref.~\cite{Wallach}. In Wallach's setup, frame functions on a tensor product of Hilbert spaces are considered that satisfy an additivity condition with respect to orthonormal bases whose elements consist of product (i.e., unentangled) vectors. It is straightforward to show that such a frame function induces a mapping on $\bold{Proj}(A,B)$ which is normalized and locally additive, but such a frame function is not necessarily locally positive. Thus the assumptions used by Wallach do not necessarily induce Dirac measures on $\bold{Proj}(A,B)$.

As we show in the next section, there exist positive Dirac measures whose associated local-density operators as given by Theorem~\ref{MTX1} are not positive semi-definite, thus positive semi-definiteness is not a necessary condition for a local-density operator to induce a positive Dirac measure on $\bold{Proj}(A,B)$. Since every density operator $\rho_{AB}\in \mathfrak{D}(AB)$ is associated with a unique positive Dirac measure via \eqref{GLEASX87}, it follows that when restricted to $\mathfrak{Dirac}^+(AB)$, the map $f$ defined in the statement of Theorem~\ref{MTX1} does \emph{not} induce a bijection $\mathfrak{Dirac}^+(AB)\to \mathfrak{D}(AB)$ between positive Dirac measures and bipartite density operators. As such, the extension to non-positive local-density operators for an analog of Gleason's Theorem for the case of local measurements performed on possibly non-spacelike separated quantum systems is absolutely necessary, even in the case of correlations induced by positive Dirac measures. 
 
\section{Examples of Dirac measures and their local-density operators} \label{S4}

In this section, we recall from the literature several examples of Dirac measures, and we give explicit formulas for their associated local-density operators. We also give an example of canonical non-negative probabilities associated with sequential measurements which do \emph{not} induce a Dirac measure on $\bold{Proj}(A,B)$, and as such, it follows from Theorem~\ref{MTX1} that such probabilities are not encoded by a local-density operator. Several of the Dirac measures we consider here will be uniquely determined by a state $\rho\in \mathfrak{D}(A)$ together with a quantum channel $\E:\A\to \B$ governing the dynamics from $A$ to $B$ (recall that $\A$ denotes the algebra of linear operators on $\H_A$, and similarly for $\B$), which by definition is a completely positive and trace-preserving (CPTP) linear map. The local-density operators associated with such Dirac measures have previously been referred to as \emph{quantum states over time} in the literature, as they inherently encode temporal correlations between timelike separated systems $A$ and $B$ corresponding to the input and output of a quantum channel. Given a quantum channel $\E:\A\to \B$, an essential role will be played by the \define{Jamio\l kowski operator} of $\E$, which is the operator $\J[\E]\in \A\otimes \B$ given by
\[
\J[\E]=(\id_A\otimes \E)(S)\, ,
\]
where $S=\sum_{i,j}|i\>\<j|\otimes |j\>\<i|$ is the swap operator.

\bx[Joint probabilities associated with spacelike separated measurements]
Let $\rho_{AB}\in \mathfrak{D}(AB)$, and let $\mu:\bold{Proj}(A,B)\to \C$ be the map given by 
\[
\mu(P\otimes Q)=\Tr[\rho_{AB}(P\otimes Q)] \qquad \forall P\otimes Q\in \bold{Proj}(A,B)\, .
\]
Then $\mu$ is a positive Dirac measure. The standard operational interpretation of $\mu$ is that it is the joint probability measure associated with measurements performed in parallel on spacelike separated systems $A$ and $B$, with the composite system $AB$ in state $\rho_{AB}$ prior to measurement. Interestingly, it was shown in Ref.~\cite{song23} that there exist density operators $\rho_{AB}$ whose correlations may also be realized by \emph{sequential} measurements performed on the same quantum system at times $t_A$ and $t_B>t_A$.
\ex

\bx[Joint probabilities associated with sequential measurements] \label{LvN}
The \define{L\"{u}ders-von~Neumann distribution} with respect to a state $\rho\in \mathfrak{D}(A)$ and a quantum channel $\mathcal{E}:\A\to \B$ is the function $\chi:\bold{Proj}(A,B)\to [0,1]$ given by
\be \label{LvNXT37}
\chi(P\otimes Q)=\Tr[\E(P\rho P)Q] \qquad \forall P\otimes Q\in \bold{Proj}(A,B)\, .
\ee
The L\"{u}ders-von~Neumann distribution is normalized and positive, but it is not locally additive for general $(\rho,\mathcal{E})$ (cf. Ref~\cite{FullwoodWu_2025}), and thus is \emph{not} a Dirac measure in general. As for the operational interpretation of $\chi$, suppose measurements are performed according to the following protocol:
\begin{itemize}
\item
System $A$ is prepared in state $\rho$.
\item
A projective measurement $\{P_i\}$ is then performed on system $A$.
\item
System $A$ then evolves according to the quantum channel $\E:\A\to \B$.
\item
A projective measurement $\{Q_j\}$ is then performed on system $B$.
\end{itemize}  
Then $\chi(P_i\otimes Q_j)$ is the joint probability of the measurement outcome $P_i$ \emph{followed by} $Q_j$ in such a 2-point sequential measurement scenario. As $\chi$ is not locally additive for general $(\rho,\mathcal{E})$, it follows from Theorem~\ref{MTX1} that for general $(\rho,\mathcal{E})$ there does not exist a local-density operator encoding the joint probabilities $\chi(P_i\otimes Q_j)$ as $\{P_i\}$ and $\{Q_j\}$ vary across all projective measurements.

If $(\rho,\mathcal{E})$ is such that $\rho$ is the maximally mixed state or if $\mathcal{E}$ is a discard-and-prepare channel---i.e., if there exists $\sigma\in \mathfrak{D}(B)$ such that $\mathcal{E}(\rho)=\sigma$ for all $\rho\in \mathfrak{D}(A)$---then it immediately follows that $\chi$ is in fact locally additive if either of these two extreme conditions hold for $(\rho,\mathcal{E})$. For $\rho$ maximally mixed the associated local-density operator is then $\mathscr{J}[\mathcal{E}]/\text{dim}(\mathcal{H}_A)$, and for $\mathcal{E}$ a discard-and-prepare channel the associated local-density operator is the product state $\rho\otimes \sigma$. In Ref.~\cite{FullwoodWu_2025}, it was conjectured that $\rho$ being maximally mixed or $\mathcal{E}$ being a discard-and-prepare channel is also a necessary condition for the local additivity of $\chi$, which was answered in the affirmative in Ref.~\cite{FY_2026}.
\ex

\bx
The \define{Kirkwood-Dirac distribution} associated with a state $\rho\in \mathfrak{D}(A)$ and a quantum channel $\mathcal{E}:\A\to \B$ is the Dirac measure $\mu:\bold{Proj}(A,B)\to \C$ given by
\[
\mu(P\otimes Q)=\Tr[\E(\rho P)Q] \qquad \forall P\otimes Q\in \bold{Proj}(A,B)\, .
\]
The local-density operator $\varrho_{AB}$ associated with the Kirkwood-Dirac distribution is given by
\[
\varrho_{AB}=\J[\mathcal{E}](\rho\otimes \mathds{1}_B)\, ,
\] 
which for general $(\rho,\E)$ is a non-Hermitian operator. The Kirkwood-Dirac distribution was introduced independently by Kirkwood in 1933~\cite{Kirkwood_1933} and Dirac in 1945~\cite{Dirac_1945} for the purpose of assigning well-defined joint probabilities to non-commuting observables, and has since come to play a prominent role in many areas of quantum physics, such as quantum thermodynamics, weak values, out-of-time-ordered correlators, quantum metrology, and support uncertainty~\cite{Alla_2014,Lost_2018,YungerHalpern_2018,Alonso_2019,Levy_2020,Lostaglio_2023,Arvidsson_2024,Gherardini_2024,Aharonov_1988,ADYLLBL20,LYABPSH22,Bievre21}. More recently, extensions of the Kirkwood-Dirac distribution to temporal, multi-time, and general dynamical processes---and their connections to temporal and spatiotemporal quantum states---have been actively investigated~\cite{Lostaglio_2023,Gherardini_2024,Schmid_2024,JiaModiKaszlikowski_2026,JiaXiao_2026}. In particular, Refs.~\cite{JiaModiKaszlikowski_2026,JiaXiao_2026} develop a multi-time temporal and spatiotemporal Kirkwood-Dirac phase-space framework that unifies a broad class of temporal state formalisms and provides direct operational access via temporal Bloch tomography. Various experimental protocols have also been established for direct experimental verification of the Kirkwood-Dirac distribution~\cite{BDOV13,Hernandez_2024,Hernandez_2024b,Wang_2024}, and recently it has been shown in Ref.~\cite{Lie_2025} that the local-density operator $\varrho_{AB}$ also admits an operational interpretation in terms of encoding the interference term in the probabilities associated with interferometric experiments.
\ex

\bx
The \define{Leifer-Spekkens} distribution associated with a state $\rho\in \mathfrak{D}(A)$ and a quantum channel $\mathcal{E}:\A\to \B$ is the function $\mu:\bold{Proj}(A,B)\to \C$ given by
\[
\mu(P\otimes Q)=\Tr[\E(\sqrt{\rho}P\sqrt{\rho})Q] \qquad \forall P\otimes Q\in \bold{Proj}(A,B)\, .
\]
The Leifer-Spekkens distribution is a positive Dirac measure, whose associated local-density operator $\varrho_{AB}$ is given by
\[
\varrho_{AB}=(\sqrt{\rho}\otimes \mathds{1}_B)\J[\mathcal{E}](\sqrt{\rho}\otimes \mathds{1}_B)\, .
\] 
We note that although the Leifer-Spekkens distribution is a positive Dirac measure, its associated local-density operator $\varrho_{AB}$ is a Hermitian operator which is \emph{not} positive semi-definite in general. As such, positive semi-definiteness of a local-density operator is not a necessary condition for its associated Dirac measure to be positive. The operational interpretation of the Leifer-Spekkens distribution is that it corresponds to the joint probabilities associated with prepare-evolve-measure scenarios. In particular, suppose $\{P_i\}$ is a PVM on system $A$. As shown in Ref.~\cite{Le06}, such a PVM uniquely determines an ensemble decomposition $\rho=\sum_i p_i \rho_i$, where $p_i=\Tr[\rho P_i]$ and (for $p_i>0$)
\[
\rho_i=\frac{\sqrt{\rho}\,P_i\sqrt{\rho}}{p_i}\, .
\]
Now suppose Alice prepares the state $\rho_i$ with probability $p_i$, and then sends the state to Bob through the quantum channel $\mathcal{E}$. Assume that Bob then performs a projective measurement $\{Q_j\}$ on the output state $\mathcal{E}(\rho_i)$. The joint probability $\mathbb{P}(i,j)$ that Alice sends state $\rho_i$ to Bob and that Bob's measurement outcome is $Q_j$ is then given by 
\[
\mathbb{P}(i,j)=p_i\Tr[\E(\rho_i)Q_j]=\Tr[\E(\sqrt{\rho}\,P_i\sqrt{\rho})Q_j]\, ,
\]
thus providing a precise operational meaning for the Dirac measure $\mu$ corresponding to the Leifer-Spekkens distribution.
\ex

\bx \label{MH}
The \define{Margenau-Hill distribution} associated with a state $\rho\in \mathfrak{D}(A)$ and a quantum channel $\mathcal{E}:\A\to \B$ is the function $\mu:\bold{Proj}(A,B)\to \C$ given by
\be \label{MHDSX71}
\mu(P\otimes Q)=\frac{1}{2}\Tr[\E(\{\rho,P\})Q] \qquad \forall P\otimes Q\in \bold{Proj}(A,B)\, ,
\ee
where $\{\cdot\, , \cdot\}$ denotes the anti-commutator. The Margenau-Hill distribution is a Dirac measure, whose associated local-density operator is the Hermitian operator $\varrho_{AB}$ given by
\be \label{CSXT71}
\varrho_{AB}=\frac{1}{2}\Big\{\rho\otimes \mathds{1}_B\, ,\J[\mathcal{E}]\Big\}\, .
\ee
The operator $\varrho_{AB}$ associated with the Margenau-Hill distribution has been referred to as the \emph{canonical state over time}, as it has been uniquely derived from different sets of assumptions in Refs.~\cite{LiNg23,PFBC23,FuPa24}. In the case that the systems $A$ and $B$ each consist of a finite number of qubits, it was shown in Refs.~\cite{HHPBS17,Liu_2025,Fullwood_2025a} that $\varrho_{AB}$ coincides with the \emph{pseudo-density operator} associated with $(\rho,\E)$. Pseudo-density operators were first introduced by Fitzsimons, Jones and Vedral for characterizing quantum correlations which imply causation~\cite{FJV15}, and they have since found applications in a wide-ranging number of topics~\cite{HLiu_2025,Marletto_2020,MVVAPGDG21,Pisar19,FuPa22a,song23,Liu_2024,Liu_2025}, including experimental verification of quantum temporal correlations, the black hole information problem, temporal quantum  teleportation, quantum channel capacity, quantum Bayesian inference, and quantum causal inference. 

As for the operational interpretation of the Margenau-Hill distribution, we first note that it is the real part of the Kirkwood-Dirac distribution, and an interferometric realization of such quasi-probabilities via quantum operations followed by post-processing was first proposed in Ref.~\cite{BDOV13}. Moreover, Johansen shows in Ref.~\cite{Johansen_2007} that the difference
\[
\mu(P\otimes Q)-\chi(P\otimes Q)
\]
admits a precise operational meaning as a measure of state disturbance associated with a two-point sequential measurement scenario whose joint probabilities correspond to the L\"{u}ders-von~Neumann distribution $\chi$. As such, the Margenau-Hill distribution may be viewed as the L\"{u}ders-von~Neumann distribution corrected by a measure of state disturbance associated with the initial measurement of a two-point sequential measurement scenario.
\ex

\bx
In Ref.~\cite{song25}, Song and Parzygnat prove a result which yields an explicit test for whether or not a local-density operator $\varrho_{AB}$ is of the form \eqref{CSXT71} for some $(\rho,\mathcal{E})$. In particular, when $\rho_A=\Tr_B[\varrho_{AB}]$ is faithful, there is a dephasing map $\mathcal{D}$ associated with an expansion of $\rho_A$ with respect to an eigenbasis $\{|i\>\}$ of $\rho_A$, and it is shown in Ref.~\cite{song25} that $\varrho_{AB}$ is of the form \eqref{CSXT71} if and only if the partial transpose of $(\mathcal{D}\otimes \id_B)(\varrho_{AB})$ with respect to the basis $\{|i\>\}$ is positive semi-definite. Now consider the one-parameter family of Hermitian local-density operators given by
\[
\varrho_{AB}(t)=\frac{1-t}{12}
\left(
\begin{array}{cccc}
-6 & \sqrt{5} & \sqrt{5} & 0 \\
\sqrt{5} & 8 & 0 & \sqrt{5} \\
\sqrt{5} & 0 & 8 & \sqrt{5} \\
0 & \sqrt{5} & \sqrt{5} & 2 \\
\end{array}
\right)+\frac{t}{4}\mathds{1}_4 \qquad t\in [0,1]\, .
\]
For all $t\in [0,1]$, the reduced density operators $\rho_A=\Tr_B[\varrho_{AB}(t)]$ and $\rho_B=\Tr_A[\varrho_{AB}(t)]$ coincide, as
\[
\rho_A=\rho_B=\frac{1-t}{6}
\left(
\begin{array}{cc}
1 & \sqrt{5} \\
\sqrt{5} & 5 \\
\end{array}
\right)
+\frac{t}{2}\mathds{1}_2 \qquad \forall t\in [0,1]\, .
\]
When $t=1$, we simply recover the maximally mixed state $\varrho_{AB}(1)=\frac{1}{4}\mathds{1}_4$, which is both a bipartite density operator and of the form \eqref{CSXT71} (with $\mathcal{E}$ the completely depolarizing channel). However, for $t$ below the threshold $t_+\approx 0.692$---specifically $t_+ = -\lambda_{\min}/(3-\lambda_{\min})$, where $\lambda_{\min}\approx -6.736$ is the smallest root of $\lambda^3-4\lambda^2-64\lambda+56=0$---the operator $\varrho_{AB}(t)$ has a negative eigenvalue and is therefore non-positive. In fact, for all $t\in [0,2/3)$, the diagonal entry $\Tr[\varrho_{AB}(t)(|0\>\<0|\otimes |0\>\<0|)]=(3t-2)/4$ is strictly negative, so the induced Dirac measure is not positive and its correlations cannot be realized by spacelike separated systems. Furthermore, applying the criterion from Ref.~\cite{song25} in the eigenbasis of $\rho_A$ shows that $\varrho_{AB}(t)$ fails to be of the form \eqref{CSXT71} for any quantum channel $\mathcal{E}$ precisely on the interval $t\in [0,t_c)$ with $t_c\approx 0.689$ (at $t=0$, $\rho_A$ is pure while $\varrho_{AB}(0)$ has non-vanishing support orthogonal to the support of $\rho_A\otimes \mathds{1}_B$, and for $t\in (0,t_c)$, the unique HPTP map solving \eqref{CSXT71} fails to be completely positive). In Ref.~\cite{FuPa23}, this family was also shown to violate subadditivity for the generalized von~Neumann entropy functional appearing in Refs.~\cite{SSW14,TTC22,CiKu23,FullwoodWu_2025a} for sufficiently small $t$ (namely $t\in [0,t_{\mathrm{sub}})$ with $t_{\mathrm{sub}}\approx 0.084$). Thus, for $t\in [0,t_c)$---and in particular at $t=0$---the local-density operator $\varrho_{AB}(t)$ describes identical systems $A$ and $B$ whose correlations can neither be realized by spacelike separated systems nor by a canonical state over time of the form \eqref{CSXT71}, leaving their operational meaning unclear.
\ex

\section{Bayes' rule for Dirac measures} \label{S5}

In this section, we derive the analog of Bayes' rule for Dirac measures.

\bn \label{BXSR71}
Let $\mu:\bold{Proj}(A,B)\to \C$ be a Dirac measure, let $\varrho_{AB}$ denote the local-density operator associated with $\mu$, and let $\overline{\mu}:\bold{Proj}(B,A)\to \C$ be the Dirac measure associated with the local-density operator $\varrho_{BA}=S\varrho_{AB}S$, where $S:\H_A\otimes \H_B\to \H_B\otimes \H_A$ is the swap operator. Then for all $P\in \bold{Proj}(A)$ and $Q\in \bold{Proj}(B)$,
\be \label{BXR587}
\mu(P\otimes Q)=\overline{\mu}(Q\otimes P)\, .
\ee
\en

\bprf
Indeed, for all $P\in \bold{Proj}(A)$ and $Q\in \bold{Proj}(B)$ we have
\begin{align*}
\overline{\mu}(Q\otimes P)&=\Tr[\varrho_{BA}(Q\otimes P)] && \text{by Theorem~\ref{MTX1}} \\
&=\Tr[S\varrho_{BA}(Q\otimes P)S] && \text{since $\text{Ad}_S$ is a $*$-isomorphism} \\
&=\Tr[(S\varrho_{BA}S)S(Q\otimes P)S] && \text{since $S^2=\mathds{1}_A\otimes \mathds{1}_B$} \\
&=\Tr[\varrho_{AB}(P\otimes Q)] && \text{since $\varrho_{BA}=S\varrho_{AB}S$} \\
&=\mu(P\otimes Q) \, ,&& \text{by Theorem~\ref{MTX1}}
\end{align*}
as desired.
\eprf

Given a Dirac measure $\mu$, the Dirac measure $\overline{\mu}$ as defined in the statement of Proposition~\ref{BXSR71} will be referred to as the \define{Bayesian reflection} of $\mu$. So now suppose projective measurements $\{P_i\}$ and $\{Q_j\}$ are performed on systems $A$ and $B$ respectively, let $\mu$ be a Dirac measure, and let $\overline{\mu}$ be its Bayesian reflection. We then define finite joint quasi-probability distributions $\mathbb{P}(i,j)$ and $\overline{\mathbb{P}}(j,i)$ given by
\[
\mathbb{P}(i,j)=\mu(P_i\otimes Q_j) \quad \text{and} \quad \overline{\mathbb{P}}(j,i)=\overline{\mu}(Q_j\otimes P_i)\, ,
\]
and let $\mathbb{P}(i)=\Tr[\rho_A P_i]$ and $\overline{\mathbb{P}}(j)=\Tr[\rho_B Q_j]$ be the associated marginal distributions, where $\rho_A$ and $\rho_B$ are the reduced density operators of the local-density operator $\varrho_{AB}$ associated with the Dirac measure $\mu$. Now by \eqref{BXR587} it follows that for all $i$ and $j$ we have
\[
\mathbb{P}(i,j)=\overline{\mathbb{P}}(j,i) \, ,
\]
thus whenever $\mathbb{P}(i)>0$ and $\overline{\mathbb{P}}(j)>0$, multiplying the LHS and RHS of the above equation by $\mathbb{P}(i)/\mathbb{P}(i)$ and $\overline{\mathbb{P}}(j)/\overline{\mathbb{P}}(j)$, respectively, yields
\[
\mathbb{P}(i)\mathbb{P}(j|i)=\overline{\mathbb{P}}(j)\overline{\mathbb{P}}(i|j)\, ,
\]
where $\mathbb{P}(j|i)$ and $\overline{\mathbb{P}}(i|j)$ are the conditional distributions given by
\[
\mathbb{P}(j|i)=\frac{\mathbb{P}(i,j)}{\mathbb{P}(i)} \quad \text{and} \quad \overline{\mathbb{P}}(i|j)=\frac{\overline{\mathbb{P}}(j,i)}{\overline{\mathbb{P}}(j)}\, .
\]
It then follows that
\be \label{NMXBXR69}
\mathbb{P}(j|i)=\frac{\overline{\mathbb{P}}(j)\overline{\mathbb{P}}(i|j)}{\mathbb{P}(i)}\, ,
\ee
which we refer to as the \define{Bayes' rule} associated with the Dirac measure $\mu$. The operator equation 
\be \label{QBRXS57}
\varrho_{BA}=S\varrho_{AB}S
\ee
is then referred to as the \define{quantum Bayes' rule}, as this is the equation which is ultimately responsible for the numerical Bayes' rule \eqref{NMXBXR69}. Moreover, as equation \eqref{NMXBXR69} recovers the classical Bayes' rule in the case that $\mu$ is a positive Dirac measure, the quantum Bayes' rule \eqref{QBRXS57} provides an operator-theoretic starting point for a derivation of the classical Bayes' rule. Finally, we note that the quantum Bayes' rule \eqref{QBRXS57} generalizes the quantum Bayes' rule from Ref.~\cite{FuPa22a} for quantum states over time to all local-density operators, thus opening up a new line of investigation into the study of quantum Bayesian inference.

\section{Concluding remarks}

In this work, we introduced the notion of a \emph{Dirac measure}, which generalizes a quasi-probability distribution introduced independently by Kirkwood and Dirac for the purpose of assigning well-defined probabilities to non-commuting observables. Given two quantum systems $A$ and $B$ which are possibly non-spacelike separated, a Dirac measure uniquely extends to a bilinear functional on the product space $\bold{Obs}(A)\times \bold{Obs}(B)$, which yields a correlation function associated with local measurements performed on systems $A$ and $B$. We then proved an analog of Gleason's Theorem for Dirac measures, showing that every Dirac measure $\mu$ may be associated with a unique bipartite operator $\varrho_{AB}$ which is a linear representation of the correlation function associated with the Dirac measure $\mu$. While such an operator $\varrho_{AB}$ is of unit trace, it is not Hermitian in general, let alone positive semi-definite. However, the reduced density operators of $\varrho_{AB}$ are genuine density operators representing the states of the local systems $A$ and $B$, thus we refer to such operators $\varrho_{AB}$ as \emph{local-density operators}. When $A$ and $B$ are spacelike separated, the notions of a bipartite density operator on the composite system $AB$ and a local-density operator on $AB$ in fact coincide, thus local-density operators generalize the notion of quantum state to possibly non-spacelike separated systems. Due to its simple definition, the notion of local-density operator also subsumes other extensions of the notion of density operator into the spatiotemporal domain, such as pseudo-density operators and quantum states over time. As such, local-density operators admit various operational interpretations, such as encoding the correlations between prepare-evolve-measure scenarios, sequential measurement scenarios, and the interference terms of interferometry. 

While a general Dirac measure is complex-valued, there are in fact positive Dirac measures whose associated local-density operators are not positive semi-definite (as was shown in the case of prepare-evolve-measure scenarios). It then follows that even if one is not interested in quasi-probabilities, and prefers a limitation to positive probabilities, then non-positive local-density operators will still be required for a linear representation of correlation functions associated with certain non-spacelike separated systems. Interestingly, the necessity for non-positive operators to describe quantum correlations which extend across both space and time is reminiscent of the way in which the signature of a spacetime metric picks up a minus sign as it extends across both space and time. This could either be coincidental, or a signal pointing toward a deeper connection between quantum correlations, space, and time. In any case, the local-density operator formalism provides a simple and straightforward generalization of the density operator formalism which is indifferent to any spacetime labels one may associate with local quantum systems, thus providing a rigorous foundation for a background independent formulation of quantum theory.  

\emph{Acknowledgments.---} J.F. is supported by the National Natural Science Foundation of China under Grant No.~W2632013, and by the Hainan Provincial Natural Science Foundation of China under Grant No.~126MS0010.

\addcontentsline{toc}{section}{\numberline{}Bibliography}
\bibliographystyle{quantum}
\bibliography{references}

\Addresses

\end{document}